\makeatletter
\@ifundefined{@parse@version@dash}{%
\def\@parse@version#1{\@parse@version@0#1}
\def\@parse@version@#1/#2/#3#4#5\@nil{%
\@parse@version@dash#1-#2-#3#4\@nil}
\def\@parse@version@dash#1-#2-#3#4#5\@nil{%
  \if\relax#2\relax\else#1\fi#2#3#4 }
}{}
\makeatother

\documentclass[%
 reprint,
 amsmath,amssymb,
 aps,
floatfix,
]{revtex4-2}

\usepackage{graphicx}
\usepackage{dcolumn}
\usepackage{bm}
\usepackage{pdfpages}
\usepackage{etoolbox} 
\makeatletter
\patchcmd{\@outputpage@head}{\@ifx{\LS@rot\@undefined}{}{\LS@rot}}{}{}{}
\makeatother
\usepackage{pgffor}

\usepackage{color}
\usepackage{comment}

\newcommand{\changed}[1]{#1}
\usepackage{hyperref}

\newcommand{\vect}[1]{\mbox{\boldmath $#1$}}
\newcommand{\fQS}{f_{\mathrm{QS}}}
\newcommand{\nfp}{n_{\mathrm{fp}}}

\begin{document}

\title{Magnetic fields with precise quasisymmetry for plasma confinement}
\author{Matt Landreman}
\affiliation{%
Institute for Research in Electronics and Applied Physics, University of Maryland, College Park, MD, 20742
}%
\author{Elizabeth Paul}
\affiliation{Princeton University, Princeton, NJ, 08544}

\date{\today}

\begin{abstract}
Quasisymmetry is an unusual symmetry that can be present in toroidal magnetic fields, enabling confinement of charged particles and plasma. 
Here it is shown that both quasi-axisymmetry and quasi-helical symmetry can be achieved to \changed{a much higher precision than previously thought} over a significant volume, resulting in exceptional confinement. 
For a 1 Tesla mean field far from axisymmetry (vacuum rotational transform $>$ 0.4), 
symmetry-breaking mode amplitudes throughout a volume of aspect ratio 6 can be made as small as the typical $\sim 50$ $\mu$T geomagnetic field.
\end{abstract}

\maketitle


\section{Introduction}

Symmetries have far-reaching consequences  throughout physics, and for charged particles in a magnetic field, a remarkable symmetry called quasisymmetry (QS) can provide confinement \cite{Boozer83,NuhrenbergZille,HelanderReview,Rodriguez}.
QS is relevant for particles with a small Larmor radius compared to scale lengths in a steady magnetic field $\vect{B}$, so the magnetic moment is an adiabatic invariant. 
In this case, axisymmetry (standard continuous rotation symmetry) turns out to be unnecessary for conservation of a canonical angular momentum; it is sufficient for $B=|\vect{B}|$ to have an invariant direction in \changed{certain special coordinate systems \cite{HelanderReview,Rodriguez}, including Boozer  \cite{BoozerCoordinates} or Hamada \cite{Hamada} coordinates.} This condition is QS, and when it holds, the constraint of canonical angular momentum conservation guarantees confinement of the collisionless trajectories up to a threshold energy.
This result is striking because the trajectories of bouncing particles in a 
magnetic field without continuous symmetry are otherwise typically not confined.
While 
\changed{
axisymmetric fields have QS, 
confinement in axisymmetry (e.g. tokamaks and dipoles) requires a significant electric current inside the confinement region. This follows from applying Amp\`{e}re's law to the poloidal magnetic field \cite{Wesson}, which is required to limit the cross-field drifts. 
 This electric current is hard to drive and sustain stably, so} QS without axisymmetry is ideal.
If realized in a stellarator, a nonaxisymmetric toroidal configuration of magnets, QS enables steady-state magnetic confinement of plasma.
QS has been the basis for several approaches to magnetic confinement fusion in stellarator devices \cite{Anderson, NCSX, CFQS}, and it is also being applied for confinement of pair plasmas \cite{stenson,stoneking}.

However, it is unclear how accurately QS (in the absence of  axisymmetry) can be realized. It has been conjectured (but not proven) that in the absence of axisymmetry, QS may be possible on an isolated surface but it cannot be perfectly realized throughout a volume \changed{\cite{GB1,GB2,PlunkNearAxisymmetry}}.
A number of nominally QS field configurations have been devised numerically \cite{NuhrenbergZille,NCSX,ARIESCS,Henneberg,CFQS,wistell},
and one has been realized experimentally \cite{Anderson},
 but these fields all have substantial deviations from symmetry, discussed below.
In these previous studies, QS was only one of several design objectives, which also included e.g. magnetohydrodynamic (MHD) stability, leaving open the fundamental question of how closely QS can be achieved.
In the present paper, it is shown that QS can be realized to far greater precision than in these previous examples, not only on a single surface but throughout a sizeable volume, yielding exceptional trajectory confinement.

\begin{figure}[hbt]
\includegraphics[width=\columnwidth]{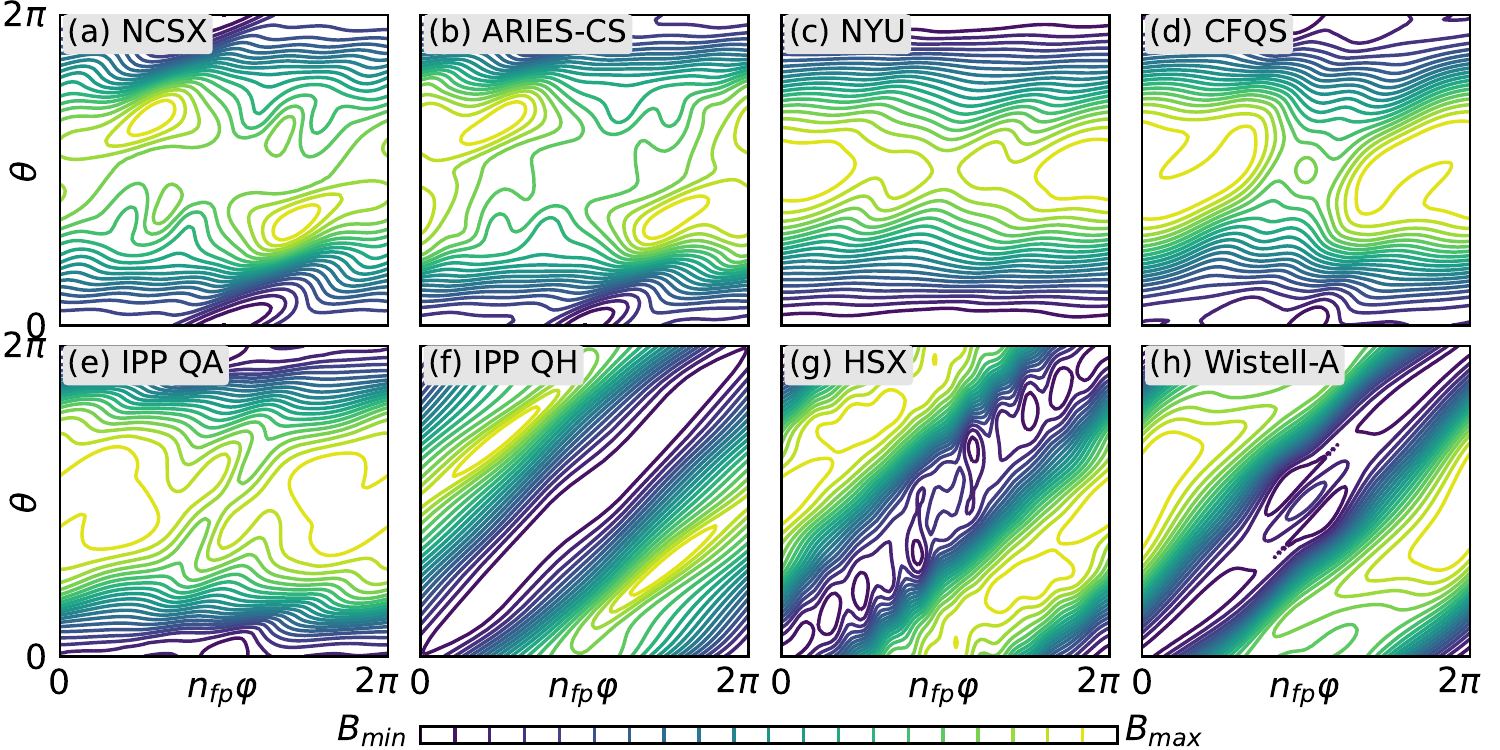}
\caption{\label{fig:previous} 
Magnetic field strength on the boundary surface of previous magnetic field configurations designed to approximate quasisymmetry,
as functions of the Boozer angles.
(a) NCSX \cite{NCSX}, (b) ARIES-CS \cite{ARIESCS}, (c) a QA developed at New York University \cite{GarabedianPNAS,GarabedianNIST}, (d) CFQS \cite{CFQS}, (e) a QA developed at the Max Planck Institute for Plasma Physics (IPP) \cite{Henneberg},
(f) a QH developed at IPP \cite{NuhrenbergZille},
(g) HSX \cite{Anderson},
(h) Wistell-A \cite{wistell}.
}
\end{figure}

Some previous examples of nominally QS fields
are shown in Fig. \ref{fig:previous}. Here and throughout this paper, fields are considered for which $\vect{B}$ is tangent to nested toroidal surfaces, ``flux surfaces''. The figure shows $B=|\vect{B}|$ on an outer flux surface for each configuration, as a function of the Boozer toroidal and poloidal angles
$(\varphi,\theta)$, angles \changed{defining} the long and short way around the torus. 
These coordinates are special because in this coordinate system the Lagrangian for drift motion varies on a flux surface only through $B$, rather than through all vector components of $\vect{B}$.
QS can be expressed as the condition $B=B(s, M \theta - N \varphi)$ for integers $M$ and $N$, so $B$ contours are straight in the $\theta$-$\varphi$ plane. Here, $s$ is the toroidal magnetic flux enclosed by a flux surface, normalized to the flux at the boundary. QS with $M \ne 1$ is impossible in the innermost region \cite{CaryShasharina}
and so will not be considered here.
QS with $N=0$ is termed quasi-axisymmetry (QA), and QS with nonzero $N$ (and $M$) is called quasi-helical (QH) symmetry. 
Figs.~\ref{fig:previous}.a-e are QA and Figs. \ref{fig:previous}.f-h are QH.
In the figure, 
quantities are periodic in $\varphi$ with period $2\pi/\nfp$ where
$\nfp$ is the number of field periods.
\changed{All configurations in Fig.~\ref{fig:previous} except HSX are fixed-boundary cases free of coil ripple.}

For all the configurations shown in Fig. \ref{fig:previous}, many $B$ contours deviate from straight lines, and some contours do not link the torus with the desired $(M,N)$ helicity. Significantly better QS has been demonstrated in a few cases, but only in a narrow volume: 
in Ref \cite{r2GarrenBoozer} (Fig. 21) for tori with aspect ratios $\ge 78$.
In the present work, precise QA and QH are demonstrated throughout significantly larger volumes (relative to the major radius.)

The new magnetic field configurations here are obtained by optimizing the shape of a boundary flux surface. 
A similar approach was used to obtain the fields in Fig. \ref{fig:previous}, though different objective functions and 
algorithms
were employed, which may partially explain the different results.
For QA, a higher aspect ratio (6.0) is considered here than in the configurations of Fig.~\ref{fig:previous}.a-e (2.6-4.5), facilitating better QS.
\changed{However the new QH shown here has an aspect ratio (8.0) within the range of the previous QH configurations in Fig.~\ref{fig:previous}.f-h (6.7-11.7).} 
It is also likely that 
QS was not obtained to the same precision in previous work because objectives besides QS were included in those optimizations as well.
Here, we focus on the fundamental question of how closely QS can be obtained in the absence of other
constraints.


\section{Methods}

To achieve QS, an objective function is minimized that includes the term
\begin{align}
\label{eq:fqs}
\fQS = \sum_{s_j} \left< 
\left( \frac{1}{B^3} \left[
(N - \iota M)\vect{B}\times\nabla B \cdot\nabla\psi 
\right.\right.\right. \hspace{0.3in} &\\
 \left.\left.\vphantom{\frac{1}{B}}\left.
-(MG+NI) \vect{B}\cdot\nabla B\right]\right)^2
\right>, & \nonumber
\end{align}
where $2\pi\psi$ is the toroidal flux, $G(\psi)$ is $\mu_0/(2\pi)$ times the poloidal current outside the surface, $I(\psi)$ is $\mu_0/(2\pi)$ times the toroidal current inside the surface,  $\iota$ is the rotational transform
and $\left< \ldots \right>$ is a flux surface average.
The sum is over a set of flux surfaces $s_j$. 
The objective (\ref{eq:fqs}) is motivated by the fact \cite{HelanderSimakovPRL} that in a QS field, 
$(\vect{B}\times\nabla B\cdot\nabla\psi)/(\vect{B}\cdot\nabla B) = (MG+NI)/(N-\iota M)$, so $\fQS=0$. The $1/B^3$ factor makes $\fQS$ dimensionless, and hence invariant if the field is scaled in length or magnitude.
Most previous QS fields have been obtained using a different
\changed{(but related \cite{RodriguezQSMeasures})}
objective based on symmetry-breaking Fourier modes of $B$, $f_{\mathrm{QS0}}=\sum_{m,n\ne N m/\nfp} (B_{m,n}/B_{0,0})^2$
where $B(s,\theta,\varphi)=\sum_{m,n}B_{m,n}(s)\cos(m\theta-\nfp n \varphi)$.
Unlike $f_{\mathrm{QS0}}$, Eq (\ref{eq:fqs})
conveniently does not require calculation of Boozer coordinates at each iteration.
Discretizing the 
flux surface average using uniform grids in 
the poloidal and toroidal angles, $\fQS$ has the form of a sum of squared residuals, so algorithms for nonlinear least-squares minimization can be applied.
Vacuum fields are considered here, so flux surface quality can be verified most straightforwardly.
A uniform grid 
$0, 0.1, \ldots, 1$
was used for $s_j$.

Other terms must be added to  (\ref{eq:fqs}) in the objective to obtain interesting solutions, at least for the initial conditions described below. For QH, we minimize $f_{\mathrm{QH}} = \fQS + (A - A_*)^2$ where $A$ is the aspect ratio of the boundary surface (as computed by the VMEC code, defined on page 12 of Ref. \cite{r2GarrenBoozer}), and $A_*$ is a specified target value. Without the aspect ratio term, $\fQS$ can be decreased to zero by increasing the aspect ratio. For QA, 
$\iota$ is bounded away from zero to avoid axisymmetric optima, using a total objective   $f_{\mathrm{QA}} = \fQS + (A - A_*)^2 + (\bar{\iota} - \bar{\iota}_*)^2$ where $\bar{\iota}=\int_0^1ds\, \iota$ and $\bar{\iota}_*$ is a specified target value. 
\changed{For the results here it was not necessary to introduce weights multiplying the $(A-A_*)$, $(\bar\iota-\bar\iota_*)$, or $\fQS$ terms.}

The independent variables for optimization are
$\{R_{m,n},\,Z_{m,n}\}$, which define the boundary surface via
%
\begin{eqnarray}
R(\theta,\phi) &=& \sum_{m,n}R_{m,n}\cos(m\vartheta-\nfp n\phi), \\ Z(\theta,\phi) &=& \sum_{m,n}Z_{m,n}\sin(m\vartheta-\nfp n\phi) , \nonumber  
\end{eqnarray}
where stellarator symmetry has been assumed, $\phi$ is the standard azimuthal angle, and $\vartheta$ is any poloidal angle. The mode $R_{0,0}$ is excluded from the parameter space to fix the spatial scale. As in \cite{simsoptVmecSpec}, the parameter space is expanded in a series of 5 steps, with modes $|m|,|n| \le j$ optimized in step $j$. For optimizing QA, the initial condition is axisymmetric with circular cross section. For optimizing QH, $R_{0,1}$ and $Z_{0,1}$ are also initialized with small nonzero values.

The optimization is carried
out using the SIMSOPT software framework \cite{SIMSOPT,SIMSOPT_Repo}, using the default
algorithm in scipy \cite{scipy} for nonlinear least-squares minimization (trust region reflective).
At each iteration, the magnetic field inside the boundary is computed using the VMEC code \cite{VMEC1983}. Gradients are computed using finite differences, with MPI for concurrent function evaluations. At the end of the optimization, the field is also computed with the SPEC code \cite{SPEC,SPEC2} to confirm the field from VMEC and verify there are no \changed{visible} magnetic islands,
\changed{shown by Poincare plots in the Supplemental Material}.
Data for the magnetic configurations are available at \cite{zenodo}.


\section{Results}

\begin{figure}[hbt]
\includegraphics[width=0.51\columnwidth]{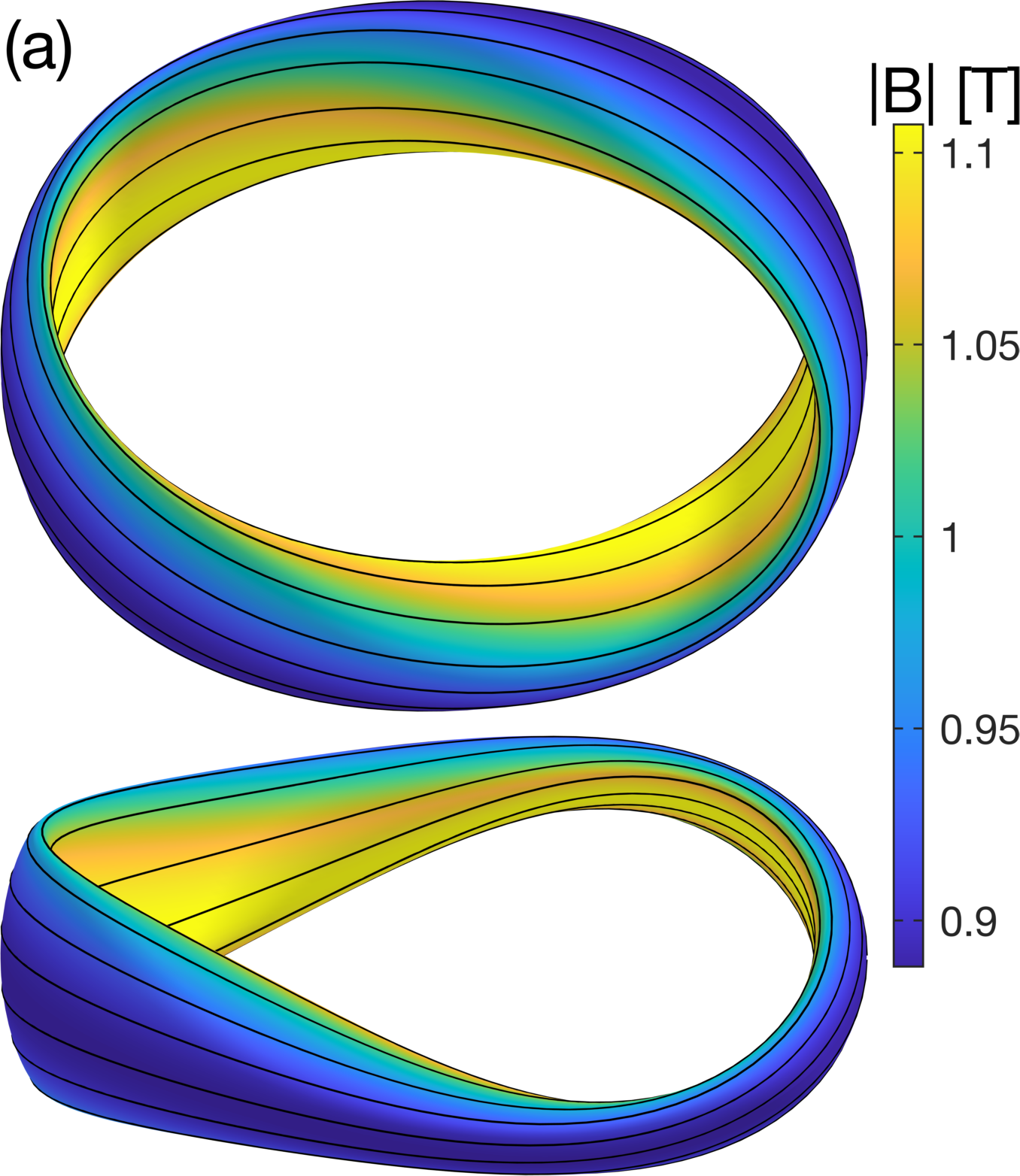}
\includegraphics[width=0.47\columnwidth]{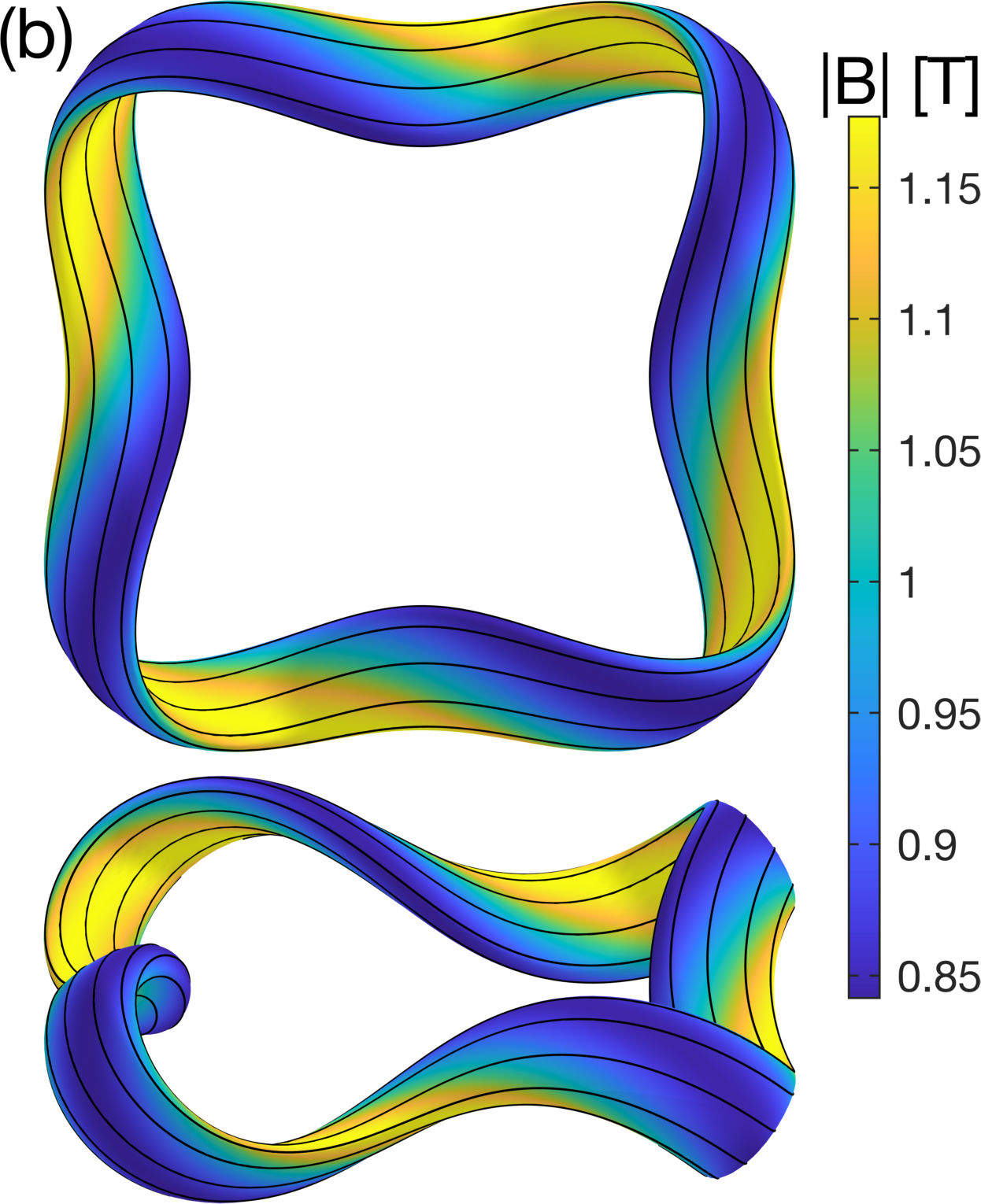}
\caption{\label{fig:3d} 
The magnetic field configurations with precise (a) quasi-axisymmetry and (b) quasi-helical symmetry, each viewed from two angles. Black curves are field lines.
}
\end{figure}

\begin{figure}[hbt]
\includegraphics[width=\columnwidth]{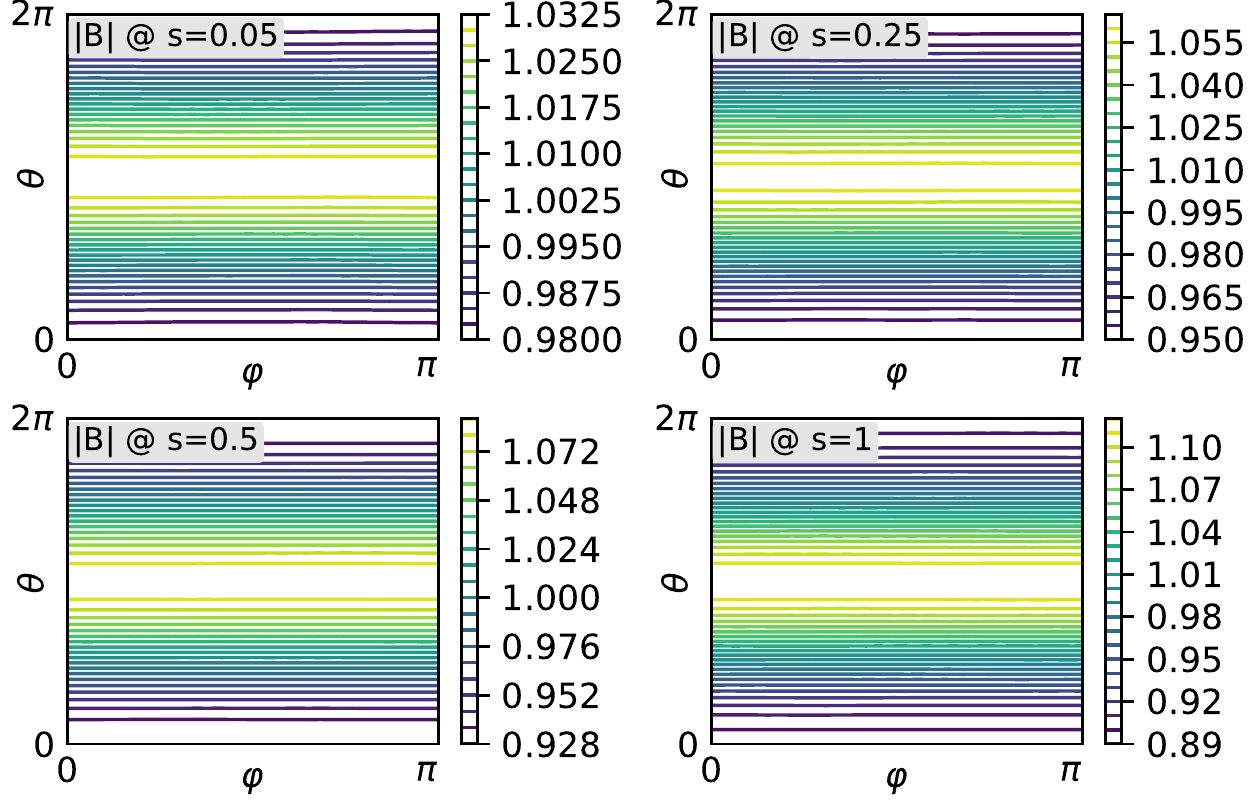}
\caption{\label{fig:QA_surfaces} 
Field strength $B$ [Tesla] on flux surfaces of the new QA field.
The contours are straight in the $\theta$-$\varphi$ plane (Boozer coordinates), demonstrating quasisymmetry.
}
\end{figure}

\begin{figure}[hbt]
\includegraphics[width=\columnwidth]{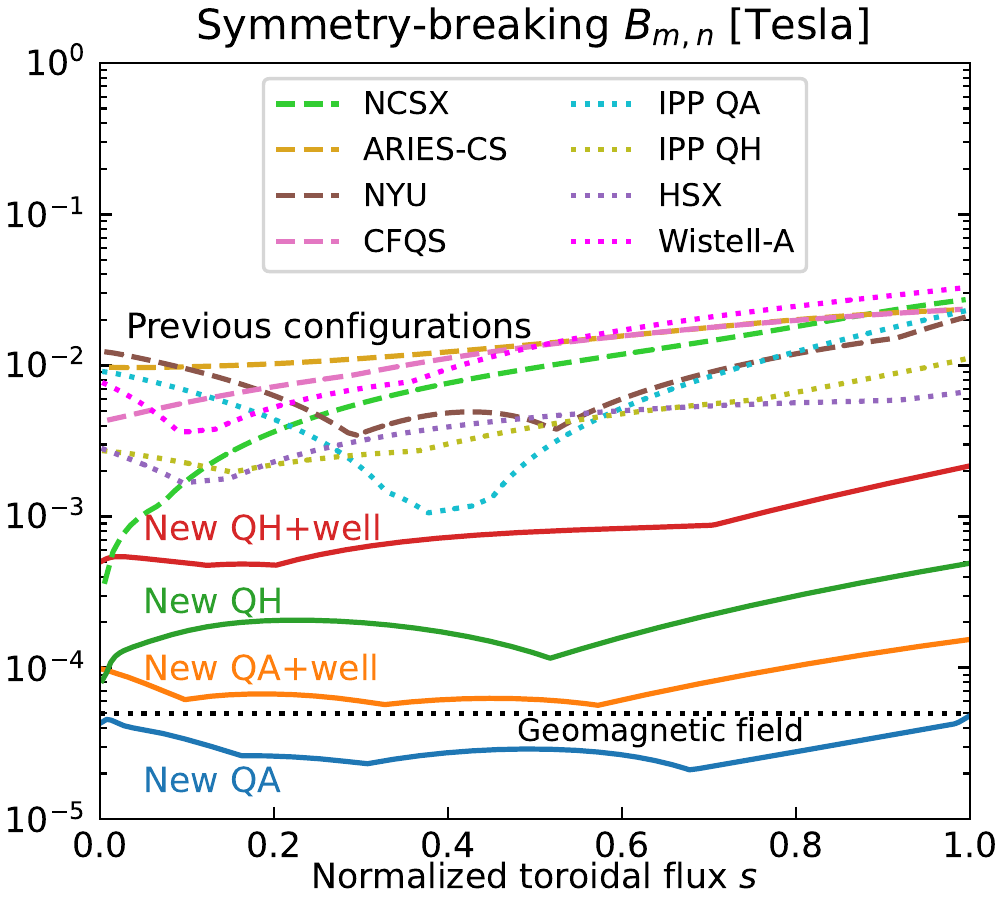}
\caption{\label{fig:compareConfigs} 
The maximum quasisymmetry-breaking mode amplitude of $B$ at each flux surface, for configurations scaled to 1 T mean $B$.
}
\end{figure}

An example of precise QA is shown in Figs.~\ref{fig:3d}-\ref{fig:QA_surfaces}. For this optimization, $\nfp=2$, $A_*=6$, and $\iota_*=0.42$ to ensure a large departure from axisymmetry, while avoiding possible magnetic islands at $\iota=2/5$.
In the optimized configuration,
\changed{$A=6.0$, and}
$\iota$ ranges from 0.423 on the magnetic axis to 0.416 at the edge, avoiding low-order rationals. 
Contours of $B$ on four surfaces are shown in Fig. \ref{fig:QA_surfaces}. Compared to Fig. \ref{fig:previous}, the contours are far straighter. The small deviations from symmetry are shown quantitatively in Fig. \ref{fig:compareConfigs}, and compared to the previous configurations of Fig.~\ref{fig:previous}.
Each configuration is scaled so the mean field $B_{0,0}$ is 1 T, matching the value for the first two QS stellarators to be built \cite{Anderson,CFQS}.
Fourier amplitudes $B_{m,n}$ that break the symmetry (those with $n \ne 0$) have amplitudes $\le 50$ $\mu$T. This value was confirmed by an independent calculation from the SPEC solution, and is comparable to the magnetic field of the Earth. 
This is the first time intrinsic QS errors have been reduced to the level of this extrinsic source of error field in a volume with experimentally relevant aspect ratio.

\begin{figure}[hbt]
\includegraphics[width=\columnwidth]{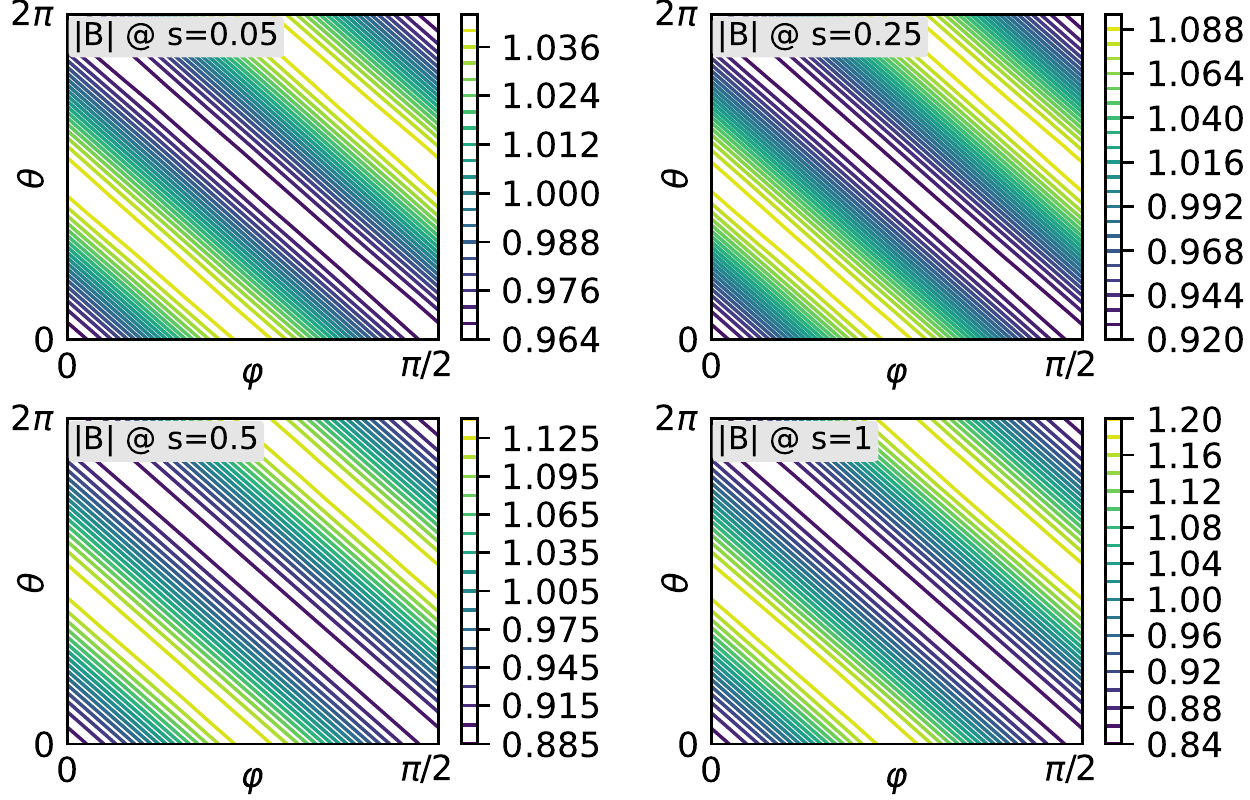}
\caption{\label{fig:QH_surfaces} $B$ [Tesla] on flux surfaces of the new QH field.
}
\end{figure}

An example with QH symmetry is shown in Figs.~\ref{fig:3d} and \ref{fig:QH_surfaces}. For this optimization, $\nfp=4$ and $A_*=8$,
\changed{with $A=8.0$ achieved.} The rotational transform is nearly constant at $\iota=1.24$.
Fig.~\ref{fig:QH_surfaces} shows that far straighter $B$ contours are possible at this $A$ than in the earlier configurations of Fig.~\ref{fig:previous}.
Errors in QS are somewhat larger than for quasi-axisymmetry at the same aspect ratio. Nonetheless, Fig.~\ref{fig:compareConfigs} shows that the $B_{m,n}$ errors in the new QH field are smaller at all radii than in any of the previous configurations, by over an order of magnitude at the boundary. 

\changed{Similar optimizations that also include magnetic well \cite{Mercier1964,Greene}, an MHD stability constraint, are discussed in the Supplemental Material. These configurations are shown in the figures as ``New QA+well'' and ``New QH+well''. In Fig.~\ref{fig:compareConfigs}, QS imperfections increase with the magnetic well constraint, but remain small compared to previous configurations.}

\begin{figure}[hbt]
\includegraphics[width=\columnwidth]{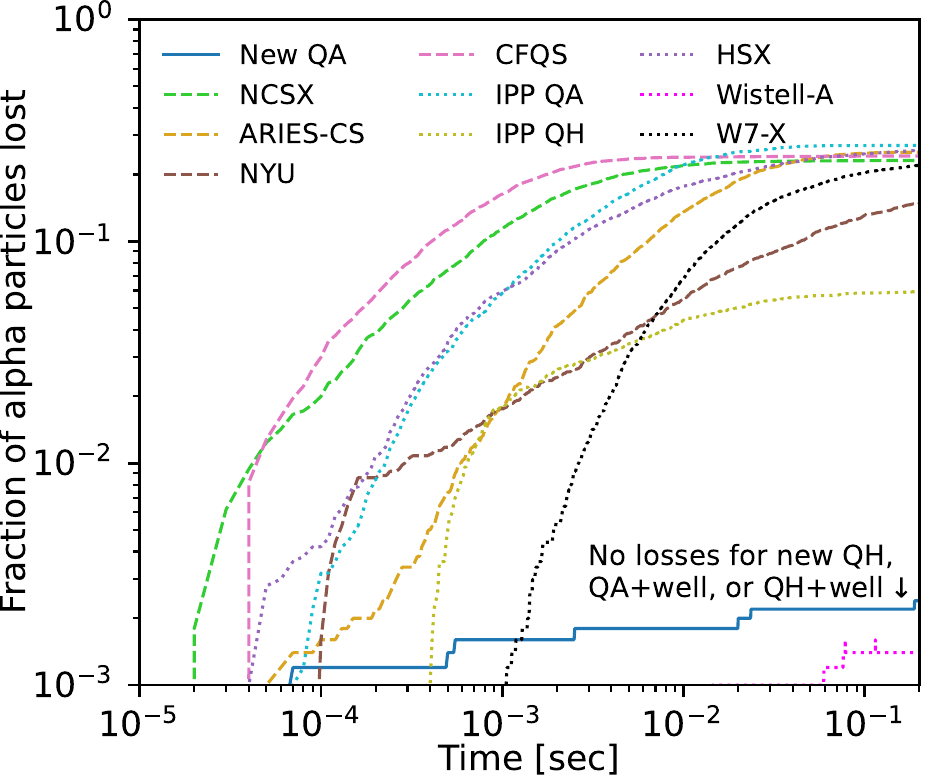}
\includegraphics[width=\columnwidth]{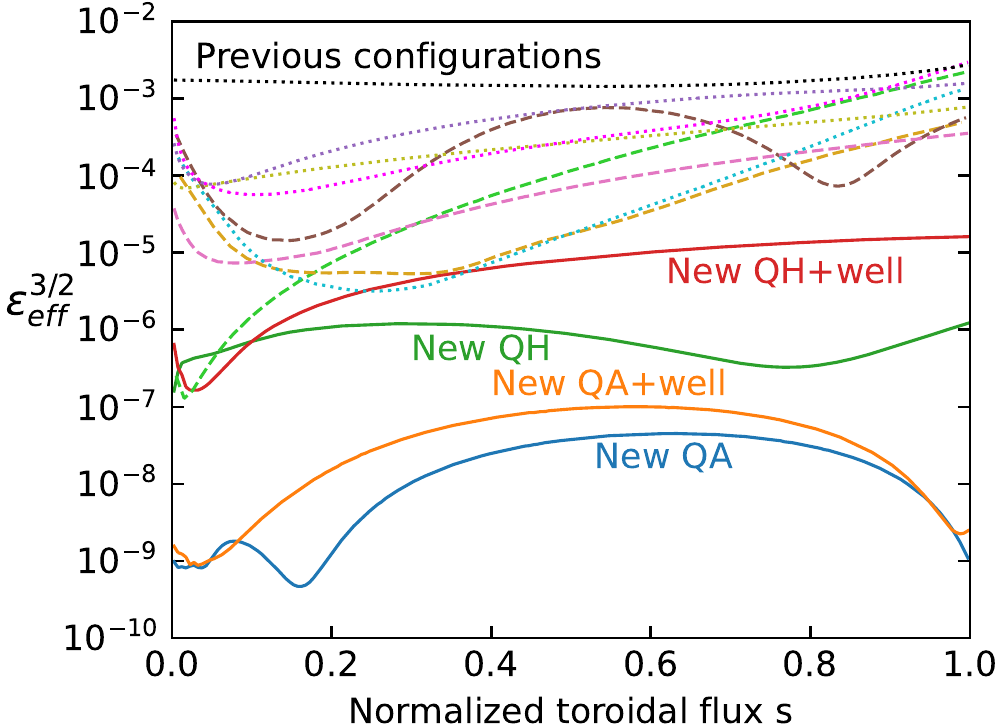}
\caption{\label{fig:losses} 
Measures of confinement.
(a) Collisionless losses of fusion-produced alpha particles initialized on the $s=0.25$ surface, in various configurations scaled to the ARIES-CS minor radius and $B$. 
(b) Collisional transport magnitude $\epsilon_{\mathrm{eff}}^{3/2}$.
}
\end{figure}

As a result of the high degree of symmetry in the new QA and QH configurations, they all exhibit excellent confinement. One measure of confinement is displayed in figure \ref{fig:losses}.a. Here, guiding center motion of test particles is followed in time. The new configurations are compared to the previous configurations of Fig.~\ref{fig:previous}, and to the largest stellarator experiment, W7-X \cite{w7x}. (For the latter, a \changed{$\beta=4$\% configuration without coil ripple is used, to give the best possible confinement.}) All configurations are scaled to the same minor radius 1.7 m and field strength $B_{0,0}(s=0)=5.7$ T of the ARIES-CS reactor \cite{ARIESCS}. An ensemble of 5000 alpha particles with 3.5 MeV, as would be produced by deuterium-tritium fusion, is initialized isotropically on the $s=0.25$ surface, and followed using 
the code of Refs.~\cite{AlbertJPP,AlbertJCP}. Particles are followed for 0.2 s, typical for the collisional slowing down time in a magnetic fusion reactor, or until they cross the $s=1$ surface and are considered lost. 
In perfect QS, there would be no losses, as long as banana orbits \changed{(the shape of trapped trajectories projected to the poloidal plane)} were sufficiently thin to stay within $s<1$.
Similar findings to Fig.~\ref{fig:losses}.a without the  new configurations were shown in Ref. \cite{BaderIAEA}.
Many of the previous configurations lose $\sim$1/4 of the particles, corresponding to those that bounce. 
The new QH, \changed{QA+well, and QH+well configurations}  perform best, with no particles lost. Though the new QA configuration \changed{without well has the best} symmetry, a few particles are still lost, which upon inspection are standard banana orbits that
extend all the way to $s>1$. 
The new QH configurations do not suffer from these losses due to thinner banana orbits, the width of which scales $\propto 1/|\iota-N|$. 
The low losses in Fig.~\ref{fig:losses} of Wistell-A, which included another measure of energetic particle confinement besides QH in its optimization \cite{Nemov2008,wistell}, show that moderate departures from QH  can still be compatible with low test particle losses, perhaps due to the reduced orbit width in QH.
\changed{The new QA+well has lower losses than the QA since $B$ varies less on each surface, so banana orbits are thinner.}

Due to the precise QS of the  new configurations, they also have 
superb confinement as measured by collisional transport for a thermal plasma.
The magnitude of this transport in the $1/\nu$ regime, where $\nu$ is the collision frequency, is known as $\epsilon_{\mathrm{eff}}^{3/2}$.
For perfect QS, $\epsilon_{\mathrm{eff}}^{3/2}$ would be zero.
As shown in Fig \ref{fig:losses}.b, $\epsilon_{\mathrm{eff}}^{3/2}$
computed by the NEO code \cite{NEO} for the new configurations is smaller
than for the other configurations of Fig \ref{fig:previous}. These values are so small that the collisional transport will almost certainly be weaker than turbulent transport.


\section{Discussion}

It has been shown here that magnetic fields exist for which QS is realized  much more precisely than in previously published examples, throughout a torus of typical stellarator aspect ratio.
This is shown for QA and QH in figures \ref{fig:QA_surfaces} and \ref{fig:QH_surfaces}, where the $B$ contours in Boozer coordinates are far straighter on many surfaces than in Fig.~\ref{fig:previous},
and shown quantitatively in Fig.~\ref{fig:compareConfigs}.
Lower symmetry breaking $B_{m,n}$ and $\epsilon_{\mathrm{eff}}^{3/2}$ were obtained at lower aspect ratio in the QAs, but better confinement of energetic particles was obtained in the QHs.

We cannot conclude that the  new configurations here have the lowest possible QS error of any QA or QH field. 
It is possible that using global optimization or other improvements in the numerical methods, configurations with even lower QS errors could be found. Moreover, QS-breaking errors can always be reduced further by increasing the aspect ratio \cite{GB1,GB2,r2GarrenBoozer}.

One natural question is
how accurately the fields here can be produced by practical magnets.
Recent innovations 
give hope that fields like the ones here could be produced with minimal errors from discrete field sources \cite{HelanderPermanentMagnets, ZhuTopology}.

More fundamentally, while the results here do not disprove the conjecture that QS cannot be achieved exactly throughout a volume, they do show that
one can come surprisingly close in practice.
While it did not seem possible in this study to reduce symmetry-breaking errors down to machine precision, 
the deviations from QS intrinsic in the new QA configuration are small enough to be subdominant to extrinsic sources of field error.


\begin{acknowledgments}
Support from the SIMSOPT development team is gratefully acknowledged.
Other magnetic configurations shown in the figures were generously provided by Aaron Bader, Michael Drevlak, 
Sophia Henneberg, 
Geoffrey McFadden, Shoichi Okamura, and John Schmitt.
We also gratefully acknowledge discussions with Christopher Albert, Per Helander, Rogerio Jorge, Don Spong, and Florian Wechsung.
This work was supported by a grant from the Simons Foundation (560651, ML).
\end{acknowledgments}




\bibliography{precise_quasisymmetry}

\newpage
\newpage
\onecolumngrid



\section*{Supplementary material}

# Supplemental material for "Magnetic fields with precise quasisymmetry"

Matt Landreman and Elizabeth Paul

November 23, 2021

This document provide more information about the new magnetic field optimizations for precise quasi-axisymmetry (QA) and quasi-helical (QH) symmetry.

## 1 Magnetic well

While it is of fundamental interest to know how accurately quasisymmetry can be achieved without additional constraints, it is also valuable to know the limits on quasisymmetry when other physics requirements are included. A common constraint imposed in stellarator design is magnetic well, related to MHD stability. For vacuum fields such as these, magnetic well is the condition $d^2V/d\psi^2 < 0$ where $V(\psi)$ is the volume enclosed by a flux surface with toroidal flux $2\pi\psi$. Here we present two optimized configurations with magnetic well, one QA and one QH. These configurations are shown as "QA+well" and "QH+well" in Figs. 4 and 6 of the main text.

To optimize for QA with magnetic well, we minimize the objective $f_{\mathrm{QAw}} = f_{\mathrm{QA}} + (W - W_*)^2$ where $W = [dV/ds(s=0) - dV/ds(s=1)]/[dV/ds(s=0)]$, $s$ is the toroidal flux normalized to its value at the boundary surface, and $W_*$ is a small positive target value. The QA+well configuration here was obtained using the choice $W_* = 0.005$, with $A_* = 6$ and $\bar{\iota}_* = 0.42$ as for the QA optimization without magnetic well. The same magnetic well objective is also effective for QH symmetry. However in the QH case it was found that the symmetry could be slightly improved by separately constraining the magnetic well at the core and the edge. The objective function used for the QH+well configuration was $f_{\mathrm{QHw}} = f_{\mathrm{QH}} + w[M(0.2, 0.4) - W_*]^2 + w[M(0.8, 1) - W_*]^2$, where $M(s_1, s_2) = 2[dV/ds(s_2) - dV/ds(s_1)]/((s_1 - s_2)[dV/ds(s_1) + dV/ds(s_2)])$ is a normalized average magnetic well over a radial interval $(s_1, s_2)$, and we chose $W_* = 0.005$ with weight $w = 3.75$. Aside from the extra terms in the objective function, the QA and QH optimizations with magnetic well were carried out with the identical procedure to those without well. For both the QA+well and QH+well optimizations, it was confirmed that at the end of the optimization, $d^2V/d\psi^2 < 0$ at all radii.

Three-dimensional views of the QA+well and QH+well configurations are shown in Fig. 1. The aspect ratios are 6.0 and 8.0 respectively, the same as for the configurations without well. The cross-sections of the four configurations are shown in Fig. 2. It can be seen that the magnetic well constraint leads to more elongated and concave shapes in the plane with standard toroidal angle $\phi = 0$. The rotational transform for the configurations is plotted in Fig. 3. The magnetic well constraint has little effect on $\iota$, and the magnetic shear remains strikingly small.

The field strength for the QA+well and QH+well configurations is displayed as a function of the Boozer angles for several flux surfaces in Fig. 4. Comparing this figure to Figs. 1, 3, and 5 of the main text, it can be seen that the constraint of magnetic well causes some slight curvature to appear in the $B$ contours, but overall the quasisymmetry remains excellent.

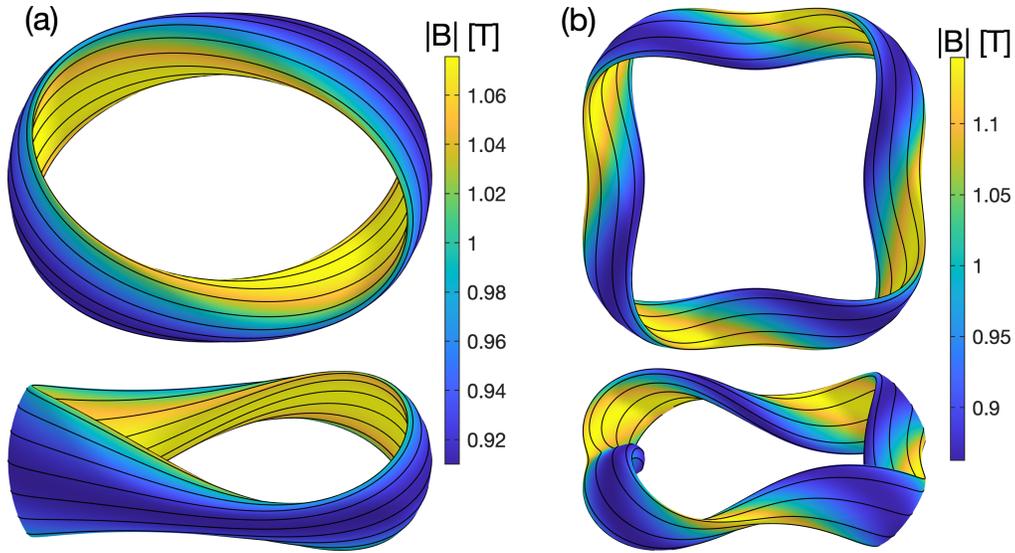

Figure 1: The new magnetic configurations with (a) precise QA and magnetic well, and (b) precise QH and magnetic well.

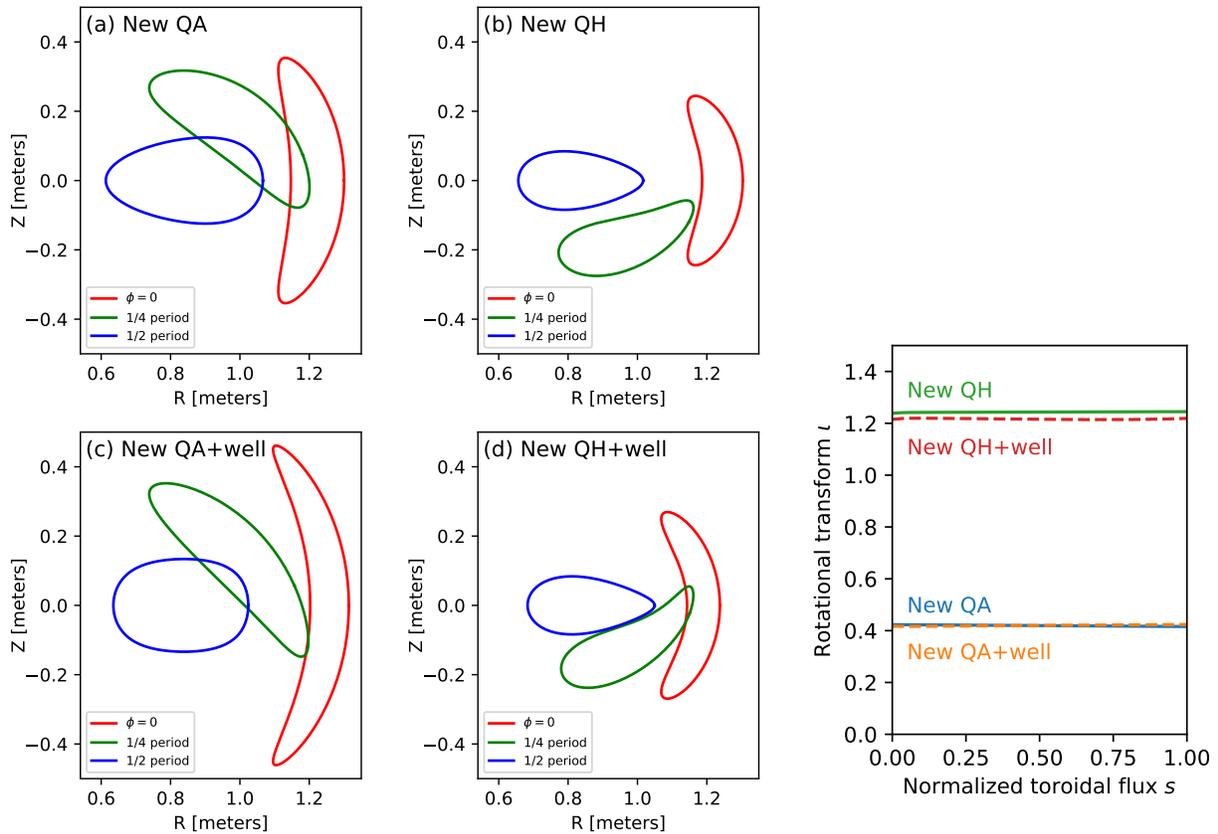

Figure 2: Cross sections of the four new configurations, each at three values of the standard toroidal angle $\phi$.

Figure 3: Radial profiles of rotational transform for the new configurations.

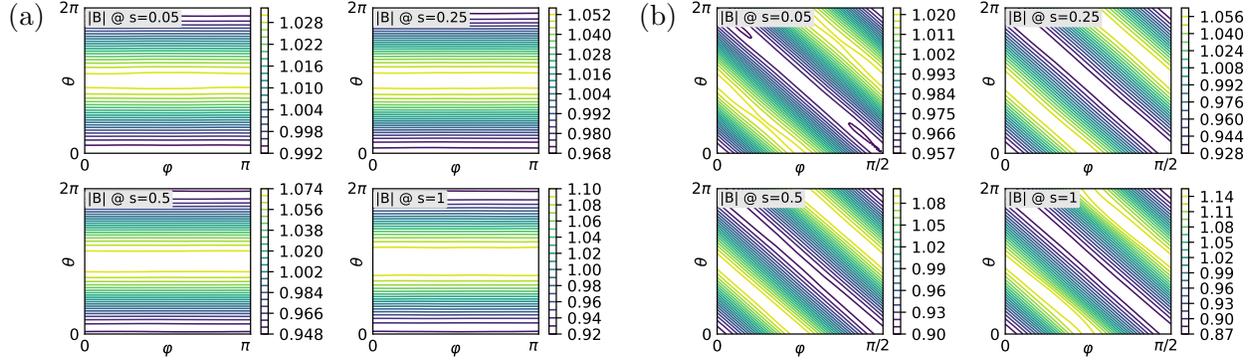

Figure 4: Magnetic field strength in Tesla for the new configurations with (a) precise QA and magnetic well, and (b) precise QH and magnetic well.

## 2 Flux surface quality

To confirm that good flux surfaces exist throughout the new QA and QH configurations, Poincare plots are shown in figure 5. Colored dots show Poincare plots generated using the SPEC code. Black curves show flux surfaces from the VMEC code, which are in good agreement with the Poincare plots.

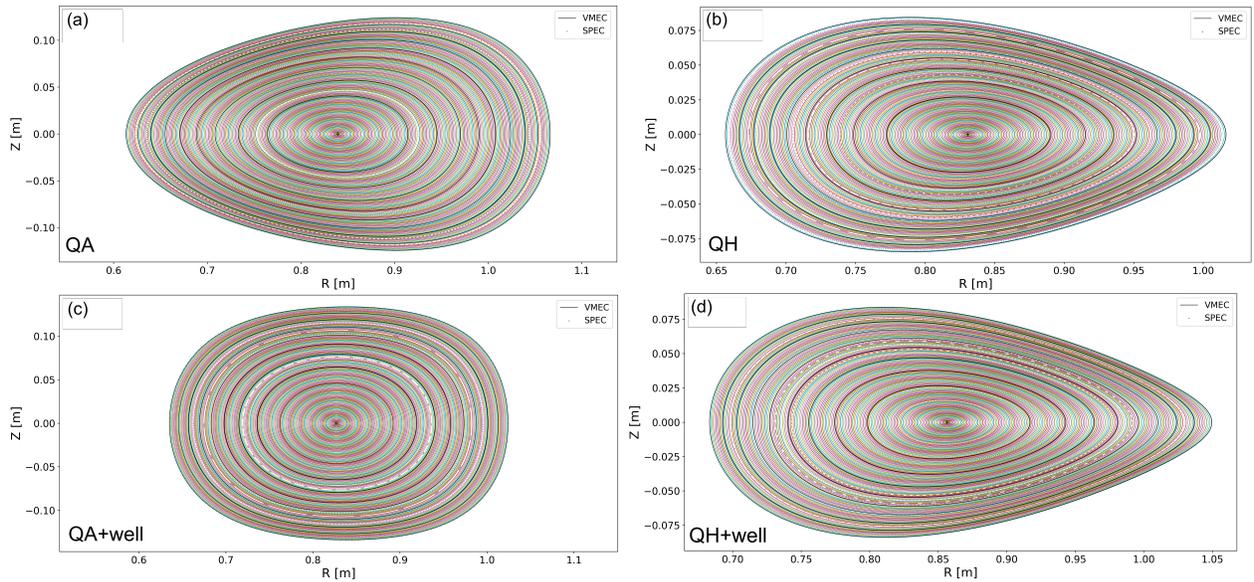

Figure 5: Verification that good flux surfaces exist throughout the new QA and QH configurations.

3



\end{document}